\documentclass[aps,prd,superscriptaddress,preprintnumbers]{revtex4}
\usepackage{graphicx,float,wrapfig,subfigure}
\usepackage{amsfonts,amsmath,amssymb,amstext}
\usepackage{latexsym}
 \usepackage{bm}
\usepackage{color}
\usepackage[normalem]{ulem}

\newcommand{\be}{\begin{equation}}
\newcommand{\ee}{\end{equation}}
\newcommand{\ba}{\begin{eqnarray}}
\newcommand{\ea}{\end{eqnarray}}

\definecolor{red}{rgb}{0.7,0,0}
\definecolor{green}{rgb}{0,0.5,0}

\begin{document}

\title{Exploring the chiral and deconfinement phase transitions in a self-consistent PNJL model}
\date{\today}

\author{Xiaozhu Yu}

\affiliation{Department of Physics, Jiangsu University, Zhenjiang 212013, P.R. China}

\author{Liangkai Wu}

\affiliation{Department of Physics, Jiangsu University, Zhenjiang 212013, P.R. China}
\author{Lang Yu}
\thanks{yulang@jlu.edu.cn}
\affiliation{College of Physics, JiLin University, Changchun 130012, P.R. China}

\author{Xinyang Wang}
\thanks{wangxy@ujs.edu.cn}
\affiliation{Department of Physics, Jiangsu University, Zhenjiang 212013, P.R. China}
\affiliation{School of Fundamental Physics and Mathematical Sciences, Hangzhou Institute for Advanced Study, UCAS, Hangzhou 310024, P.R. China}

\begin{abstract}
In this work, we study the chiral and deconfinement phase transitions in a two-flavor Polyakov loop extended Nambu--Jona-Lasinio (PNJL) model. And note that the self-consistent mean field approximation is employed by introducing an arbitrary parameter $\alpha$ to measure the weights of the Fierz-transformed interaction channels. By making use of this model, we systematically investigate the chiral and deconfinement phase transition lines (as well as the chiral ones in the NJL model for comparison) under different values of $\alpha$. It is found that, the increasing of $\alpha$ helps to enhance the chiral (pseudo)critical temperature at fixed chemical potential, and also to enhance the chiral (pseudo)critical chemical potential at fixed temperature. And the critical end point (CEP) vanishes when $\alpha$ becomes large enough. Besides, we find that the incorporation of Polyakov loop increases $T_{CEP}$ but does not change $\mu_{CEP}$ for small values of $\alpha$.
\end{abstract}
\maketitle

\section{Introduction}

Exploring the phase structure of Quantum Chromodynamics (QCD) matter is one of the most important topics in the high energy nuclear physics. With the increasing temperature or baryon density, the hadronic phase will transfer to the deconfined phase with the approximate chiral symmetry restoration, which is so-called quark-gluon plasma (QGP). It is well know now that the QGP should exist in the early universe in nature. In recent years, the heavy-ion collision experiments at the Relativistic Heavy Ion Collider (RHIC) in Brookhaven and the Large Hadron Collider (LHC) at CERN have produced the QGP at very high energies. On the other hand, the astrophysics observations could help us to have a better understanding QCD matter at the large chemical potential.

From the theoretical point of view, because the perturbation QCD method becomes unavailable in the low energy regime where the strong coupling constant is not small, we have to rely on the non-perturbation methods, such as lattice QCD.  As we know, lattice QCD calculations show that at small baryon chemical potential the chiral phase transition as well as deconfinement phase transition are crossover at finite temperature. Unfortunately, the study of QCD using lattice simulations at large baryon chemical potential is not possible at present due to the famous fermion sign problem. Therefore, the effective theories and models of QCD in the low energy regime are needed, and the QCD phase transitions have been explored by a variety of different theories and models. For instance, it is generally believed that at high baryon density the chiral phase transition is of the first order, and there exists a QCD critical end point (CEP) in the temperature and baryon chemical potential plane. The Nambu-Jona-Lasinio (NJL) model~\cite{Nambu:1961tp, Nambu:1961fr} is one of the most useful tools to study the QCD matter. The NJL model incorporates the chiral symmetry and its spontaneous breaking, the gluonic degrees of freedom are replaced by a local four-point fermion interactions. And this model has a wide range of applications for low-energy QCD research, e.g.,the weak processes in hadrons, chiral phase transition, spin structures, the form factors of the nucleons, and etc. (for reviews see Refs.~\cite{Klevansky:1992qe, Hatsuda:1994pi, Buballa:2003qv})

But since there is no gluonic degrees of freedom in the NJL model, it can not describe the color confinement of QCD. Fortunately, the decofinement phase transition could be investigated effectively in a slightly simple way that we can introduce the contribution of the Polyakov loop into the NJL model~\cite{Fukushima:2003fw}. This is so called the Polyakov loop extended Nambu--Jona-Lasinio (PNJL)  model. The essential feature of the PNJL model is that we could study both chiral phase transition and deconfinement phase transition at the same time.

 As discussed in Ref.~\cite{Klevansky:1992qe}, the Fierz transformation of the NJL interaction Lagrangian density $\mathcal{L}_F$ is actually of equal
 importance with the original Lagrangian density $\mathcal{L}$. Therefore, by inserting a new arbitrary parameter $\alpha$, a redefined Lagrangian density could be constructed by linearly combining these two types of Lagrangian densities  i.e., $\mathcal{L}_R = (1-\alpha)\mathcal{L}+\alpha\mathcal{L}_F$. Different from the conventional setting of $\alpha = 0$ in the  original NJL model and the setting of $\alpha = 1/2$ suggested in Ref.~\cite{Klevansky:1992qe}, the authors of Ref.~\cite{Wang:2019uwl} pointed out that there is no strict physical requirement that the value of $\alpha$ must be 0 or $1/2$. The weighting parameter $\alpha$ should determined by the heavy ion collision experiments or astronomy observations. Theoretically, the QCD phase diagram with different $\alpha$ have been studied by applying the self-consistent mean field approximation in the two-flavor and three-flavor NJL model~\cite{Wang:2019uwl,Yu:2020dnj}. Therefore, in order to have a better understanding of the QCD phase transitions, including both chiral phase transition and deconfinement phase transition, the method of the self-consistent mean field approximation is extended to the PNJL model. In this paper we will show how the weighting parameter $\alpha$ affects the chiral and deconfinement phase transitions of QCD matter.

This paper is organized as follows: in the next section, we apply the self-consistent mean field method on the 2-flavor PNJL model. The results of the phase transitions are discussing in section~\ref{sec-3}. The summary and discussion are given at the end of this paper.  And we assume the masses of u and d quarks are the same in this work.

\section{Self-consistent mean field approximation in the PNJL model and the gap equations}

The combination of the original Lagrangian density $\mathcal{L}$ and its Fierz transformation $\mathcal{L}_F$ in the 2-flavor NJL model is given by~\cite{Wang:2019uwl}:
\ba
\mathcal{L}_{R} = (1-\alpha)\mathcal{L}+\alpha\mathcal{L}_F,
\ea
where the quark masses $m = diag(m_u,m_d)$,
\ba
\mathcal{L}&=& \bar{\psi}(i \partial-m) \psi+ G\left[(\bar{\psi} \psi)^{2}+\left(\bar{\psi} i \gamma^{5} \cdot \tau \psi\right)^{2}\right],
\ea
and
\ba
\mathcal{L}_F&=& \bar{\psi}(i \partial-m) \psi+
\frac{G}{8 N_{c}}\left[2(\bar{\psi} \psi)^{2}+2\left(\bar{\psi} i \gamma^{5} \tau \psi\right)^{2}-2(\psi \tau \psi)^{2}\right.\nonumber\\
&&-2\left(\bar{\psi} i \gamma^{5} \psi\right)^{2}-4\left(\bar{\psi} \gamma^{\mu} \psi\right)^{2}-4\left(\bar{\psi} i \gamma^{\mu} \gamma^{5} \psi\right)^{2} 
+\left.\left(\bar{\psi} \sigma^{2 m} \psi\right)^{2}-\left(\bar{\psi} \sigma^{\mu \nu} \tau \psi\right)^{2}\right]. 
\ea

In order to study the deconfinement phase transition effectively, we add the Polyakov-loop effective potential $\mathcal{U}(\Phi, \bar{\Phi}, T)$ in the following ansatz
\ba
\frac{\mathcal{U}(\Phi, \bar{\Phi}, T)}{T^{4}}&=&-\frac{a(T)}{2} \bar{\Phi} \Phi+b(T) \ln \left[1-6 \bar{\Phi} \Phi+4\left(\bar{\Phi}^{3}+\Phi^{3}\right)-3(\bar{\Phi} \Phi)^{2}\right],
\ea
where  $a(T)=a_{0}+a_{1}\left(\frac{T_{0}}{T}\right)+a_{2}\left(\frac{T_{0}}{T}\right)^{2} \text { and } b(T)=b_{3}\left(\frac{T_{0}}{T}\right)^{3}$. In this manner, we obtain the Lagrangian density of the self consistent PNJL model
\ba
\mathcal{L}_{PNJL}&=&\bar{\psi}(i \partial-m) \psi+(1-\alpha) G\left[(\bar{\psi} \psi)^{2}+\left(\bar{\psi} i \gamma^{5} \cdot \tau \psi\right)^{2}\right]+\alpha \frac{G}{8 N_{c}}\left[2(\bar{\psi} \psi)^{2}+2\left(\bar{\psi} i \gamma^{5} \tau \psi\right)^{2}-2(\psi \tau \psi)^{2}\right.\nonumber\\
&-&\left.2\left(\bar{\psi} i \gamma^{5} \psi\right)^{2}-4\left(\bar{\psi} \gamma^{\mu} \psi\right)^{2}-4\left(\bar{\psi} i \gamma^{\mu} \gamma^{5} \psi\right)^{2} +\left(\bar{\psi} \sigma^{2 m} \psi\right)^{2}-\left(\bar{\psi} \sigma^{\mu \nu} \tau \psi\right)^{2}\right]-\mathcal{U}(\Phi, \bar{\Phi}, T).
\ea

By following the same procedure in the mean field approximation~\cite{Kapusta:2006pm}, we could write down the corresponding thermodynamic potential $\Omega_{PNJL}$,
\ba
\Omega_{PNJL}&=& \mathcal{U}(\Phi, \bar{\Phi}, T)+\frac{(M-m)^2}{4G(1-\alpha +\frac{\alpha}{4N_c})}-\frac{N_c(\mu-\tilde{\mu})^2}{2\alpha G}-2 N_{c} N_{f} \int_{0}^{\Lambda} \frac{d^{3} p}{(2 \pi)^{3}}\left[ E_{p}\right] \nonumber \\
&&-2N_f T \int \frac{d^{3} p}{(2 \pi)^{3}}\left[\ln \left(1+3 \Phi e^{-\beta\left(E_{p}-\tilde{\mu}\right)}+3 \bar{\Phi} e^{-2 \beta\left(E_{p}-\tilde{\mu}\right)}+e^{-3 \beta\left(E_{p}-\tilde{\mu}\right)}\right)\right] \nonumber\\
&&-2N_f T \int \frac{d^{3} p}{(2 \pi)^{3}}\left[\ln \left(1+3 \bar{\Phi} e^{-\beta\left(E_p+\tilde{\mu}\right)} +3 \Phi e^{-2 \beta\left(E_p+\tilde{\mu}\right)}+e^{-3 \beta\left(E_p+\tilde{\mu}\right)}\right)\right].
\ea
Here, we have $N_c = 3$ and $N_f =2$. The dispersion relation of the quark is defined by $E_p = \sqrt{\mathbf{p}^2+M^2}$, where $M$ is the constituent quark mass.
 
And then, the gap equations are determined by the stationary point of the themodynamic potential, i.e.,
\ba
\frac{\partial \Omega_{P}}{\partial M}=\frac{\partial \Omega_{P}}{\partial \Phi}=\frac{\partial \Omega_{P}}{\partial \bar{\Phi}}=\frac{\partial \Omega_{P}}{\partial \tilde{\mu}}=0.
\ea
The above oequations can be written explicitly in the following,
\begin{subequations}
\ba
\label{gap1}
M &=& m +4G(1-\alpha+\frac{\alpha}{4N_c}) N_cN_f\int_0^{\Lambda} \frac{d^3 p}{(2\pi)^3}\frac{M}{E_p}\nonumber\\
&&-4G(1-\alpha+\frac{\alpha}{4N_c}) N_f\int \frac{d^3 p}{(2\pi)^3}\frac{M}{E_p}\frac{3 \Phi e^{-\beta\left(E_{p}-\tilde{\mu}\right)}+6 \bar{\Phi} e^{-2 \beta\left(E_{p}-\tilde{\mu}\right)}+3e^{-3 \beta\left(E_{p}-\tilde{\mu}\right)} }{\left(1+3 \Phi e^{-\beta\left(E_{p}-\tilde{\mu}\right)}+3 \bar{\Phi} e^{-2 \beta\left(E_{p}-\tilde{\mu}\right)}+e^{-3 \beta\left(E_{p}-\tilde{\mu}\right)}\right)}\nonumber\\
&&-4G(1-\alpha+\frac{\alpha}{4N_c}) N_f\int \frac{d^3 p}{(2\pi)^3}\frac{M}{E_p}\frac{3 \Phi e^{-\beta\left(E_{p}+\tilde{\mu}\right)}+6 \bar{\Phi} e^{-2 \beta\left(E_{p}+\tilde{\mu}\right)}+3e^{-3 \beta\left(E_{p}+\tilde{\mu}\right)} }{\left(1+3 \Phi e^{-\beta\left(E_{p}+\tilde{\mu}\right)}+3 \bar{\Phi} e^{-2 \beta\left(E_{p}+\tilde{\mu}\right)}+e^{-3 \beta\left(E_{p}+\tilde{\mu}\right)}\right)},
\ea
\ba
\label{gap2}
\tilde{\mu}&=&   \mu -\frac{2\alpha G}{N_c} N_f\int \frac{d^3 p}{(2\pi)^3}\frac{3 \Phi e^{-\beta\left(E_{p}-\tilde{\mu}\right)}+6 \bar{\Phi} e^{-2 \beta\left(E_{p}-\tilde{\mu}\right)}+3e^{-3 \beta\left(E_{p}-\tilde{\mu}\right)} }{\left(1+3 \Phi e^{-\beta\left(E_{p}-\tilde{\mu}\right)}+3 \bar{\Phi} e^{-2 \beta\left(E_{p}-\tilde{\mu}\right)}+e^{-3 \beta\left(E_{p}-\tilde{\mu}\right)}\right)}\nonumber\\
&&+\frac{2\alpha G}{N_c} N_f\int \frac{d^3 p}{(2\pi)^3}\frac{3 \Phi e^{-\beta\left(E_{p}+\tilde{\mu}\right)}+6 \bar{\Phi} e^{-2 \beta\left(E_{p}+\tilde{\mu}\right)}+3e^{-3 \beta\left(E_{p}+\tilde{\mu}\right)} }{\left(1+3 \Phi e^{-\beta\left(E_{p}+\tilde{\mu}\right)}+3 \bar{\Phi} e^{-2 \beta\left(E_{p}+\tilde{\mu}\right)}+e^{-3 \beta\left(E_{p}+\tilde{\mu}\right)}\right)},
\ea
\ba
\label{gap3}
0 &=&T^4\left[ -\frac{a(T)}{2}\bar{\Phi}+b(T)\frac{-6 \bar{\Phi}+ 12 \Phi^2 -6\bar{\Phi}^2\Phi}{1-6 \bar{\Phi} \Phi+4\left(\bar{\Phi}^{3}+\Phi^{3}\right)-3(\bar{\Phi} \Phi)^{2}}\right]\nonumber\\
&&-2N_f T\int \frac{d^3 p}{(2\pi)^3}\frac{3e^{- \beta\left(E_{p}-\tilde{\mu}\right)} }{\left(1+3 \Phi e^{-\beta\left(E_{p}-\tilde{\mu}\right)}+3 \bar{\Phi} e^{-2 \beta\left(E_{p}-\tilde{\mu}\right)}+e^{-3 \beta\left(E_{p}-\tilde{\mu}\right)}\right)}\nonumber\\
&&-2N_f T\int \frac{d^3 p}{(2\pi)^3}\frac{3e^{- \beta\left(E_{p}+\tilde{\mu}\right)} }{\left(1+3 \Phi e^{-\beta\left(E_{p}+\tilde{\mu}\right)}+3 \bar{\Phi} e^{-2 \beta\left(E_{p}+\tilde{\mu}\right)}+e^{-3 \beta\left(E_{p}+\tilde{\mu}\right)}\right)}
\ea
and
\ba
\label{gap4}
0 &=&T^4\left[-\frac{a(T)}{2}\Phi+b(T)\frac{-6 \Phi+ 12 \bar{\Phi}^2 -6\bar{\Phi}\Phi^2}{1-6 \bar{\Phi} \Phi+4\left(\bar{\Phi}^{3}+\Phi^{3}\right)-3(\bar{\Phi} \Phi)^{2}}\right]\nonumber\\
&&-2N_f T \int \frac{d^3 p}{(2\pi)^3}\frac{3e^{- 2\beta\left(E_{p}-\tilde{\mu}\right)} }{\left(1+3 \Phi e^{-\beta\left(E_{p}-\tilde{\mu}\right)}+3 \bar{\Phi} e^{-2 \beta\left(E_{p}-\tilde{\mu}\right)}+e^{-3 \beta\left(E_{p}-\tilde{\mu}\right)}\right)}\nonumber\\
&&-2N_f T\int \frac{d^3 p}{(2\pi)^3}\frac{3e^{- 2\beta\left(E_{p}+\tilde{\mu}\right)} }{\left(1+3 \Phi e^{-\beta\left(E_{p}+\tilde{\mu}\right)}+3 \bar{\Phi} e^{-2 \beta\left(E_{p}+\tilde{\mu}\right)}+e^{-3 \beta\left(E_{p}+\tilde{\mu}\right)}\right)}.
\ea
\end{subequations}

By fitting the empirical values of the pion mass $m_{\pi} = 138$ MeV, the pion decay constant
$f_\pi = 93$ MeV and the quark condensate per flavour $\left<\bar{\phi}\phi\right> = -(250 \text{MeV})^3$, we first fix three parameters, i.e., the current mass $m = 5.5\text{MeV}$~\cite{Hatsuda:1994pi}, the coupling constant of the conventional NJL model $g = 5.074 \times 10^{-6} \text{MeV}^{-2}$ and the three dimensional cut-off $\Lambda = 631 \text{MeV}$ . And then, we redefine the coupling constant G as Ref.~\cite{Yang:2019lyn} that,
\ba
G = \frac{1+\frac{1}{4N_c}}{1-\alpha+\frac{\alpha}{4N_c}}g.
\ea
Besides, we take $T_0=270 {\rm MeV}$ with $a_0 = 3.51$, $a_1 = -2.47$, $a_2 = 15.22$, $b_3 = -1.75$ in our model~\cite{Li:2018ygx}. Since the evaluation of the chiral condensate is our main purpose in this paper, we use the cut-off on the thermal part of the integrals in the gap equations~\cite{Xue:2021ldz}.

\section{chiral and deconfinement phase transitions}
\label{sec-3}
\begin{figure}[t]
\centering
\subfigure[]{\includegraphics[width=0.45\textwidth]{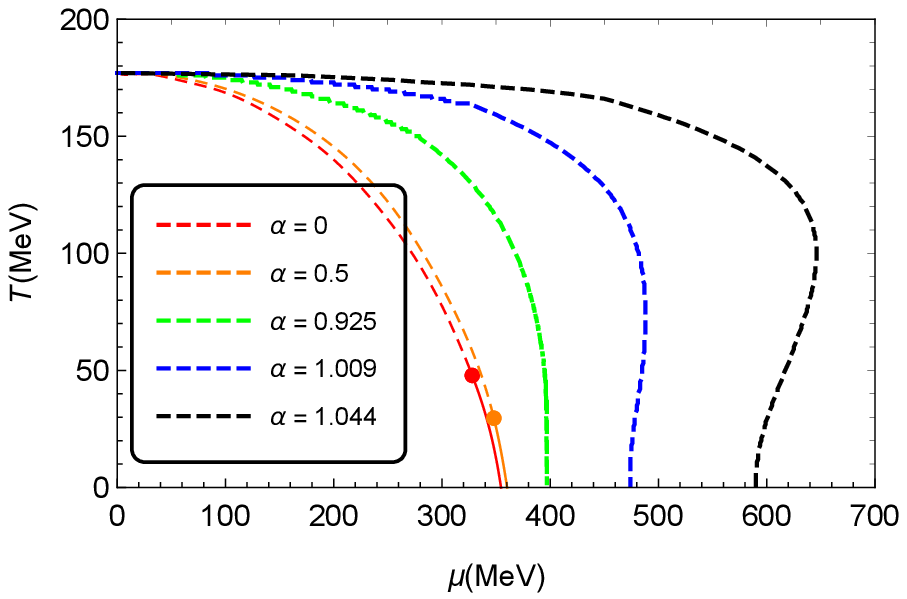}}
\hspace{0.01\textwidth}
\subfigure[]{\includegraphics[width=0.45\textwidth]{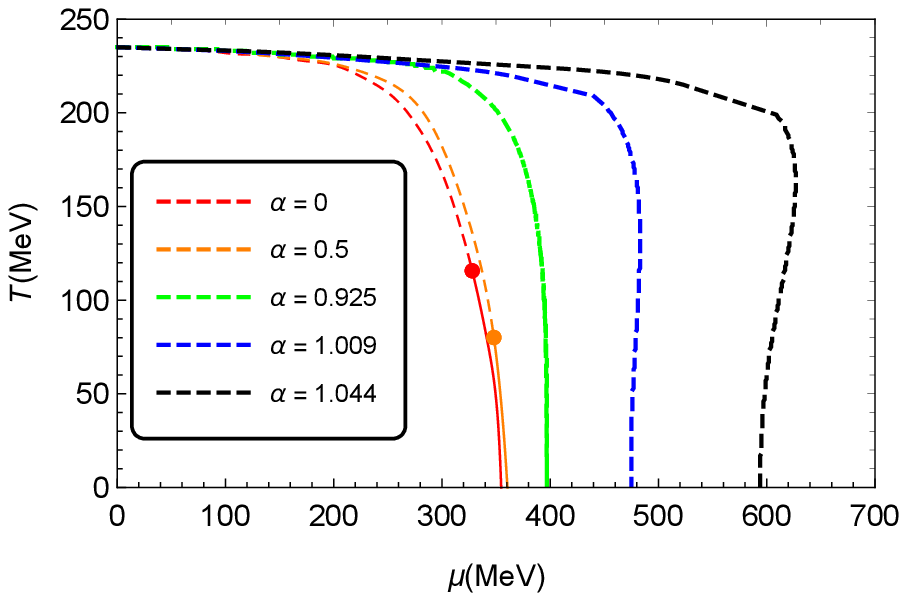}}
\caption{Chiral phase transition in NJL model(left panel) and PNJL model(right panel) with $\alpha = 0, 0.5, 0.925, 1.009, 1.044$.  The dashed line are represented the crossover transition and the bold line are represented the first order transition. The CEP for $\alpha = 0, 0.5$ are shown as big dots.}
\label{figure1}
\end{figure}

\begin{figure}[t]
\centering
\subfigure[]{\includegraphics[width=0.45\textwidth]{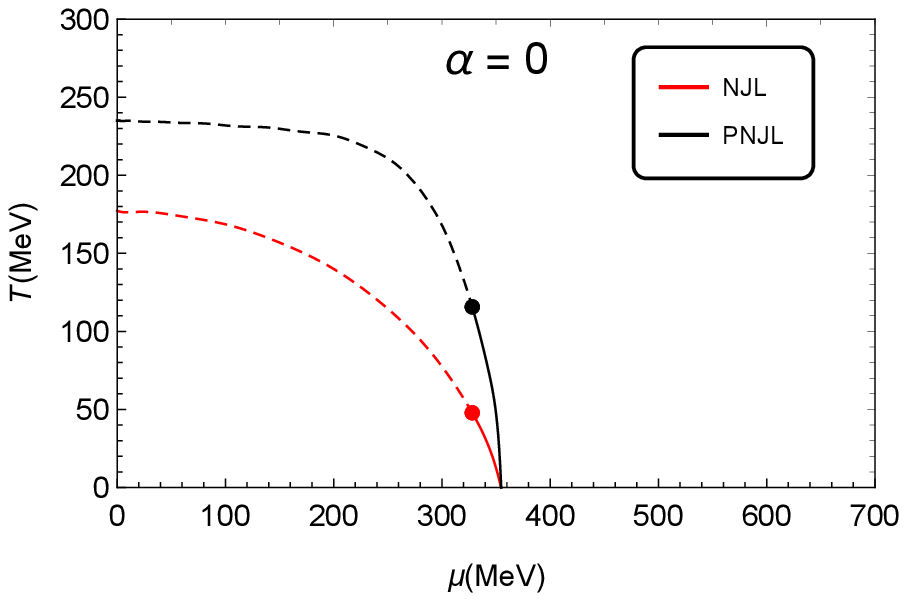}}
\hspace{0.01\textwidth}
\subfigure[]{\includegraphics[width=0.45\textwidth]{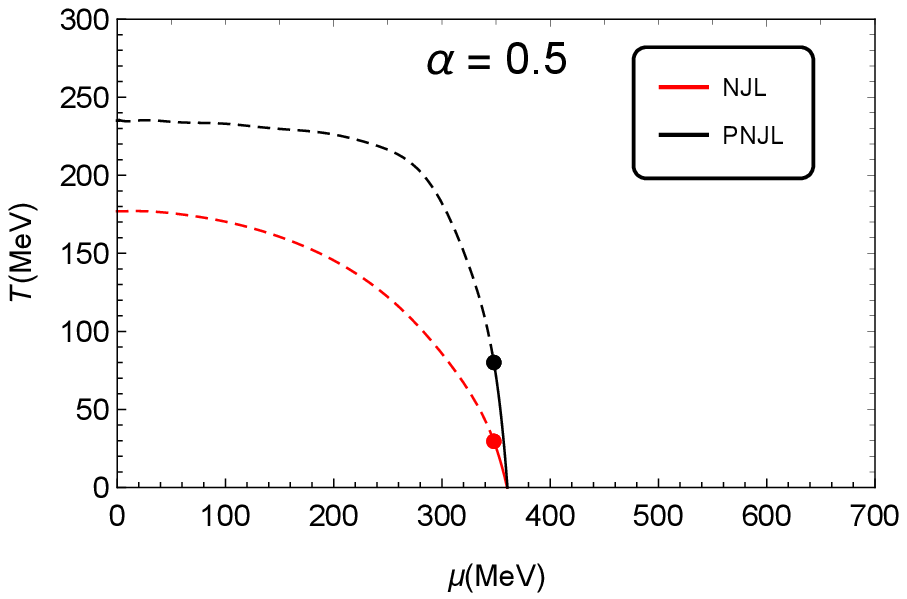}}
\subfigure[]{\includegraphics[width=0.45\textwidth]{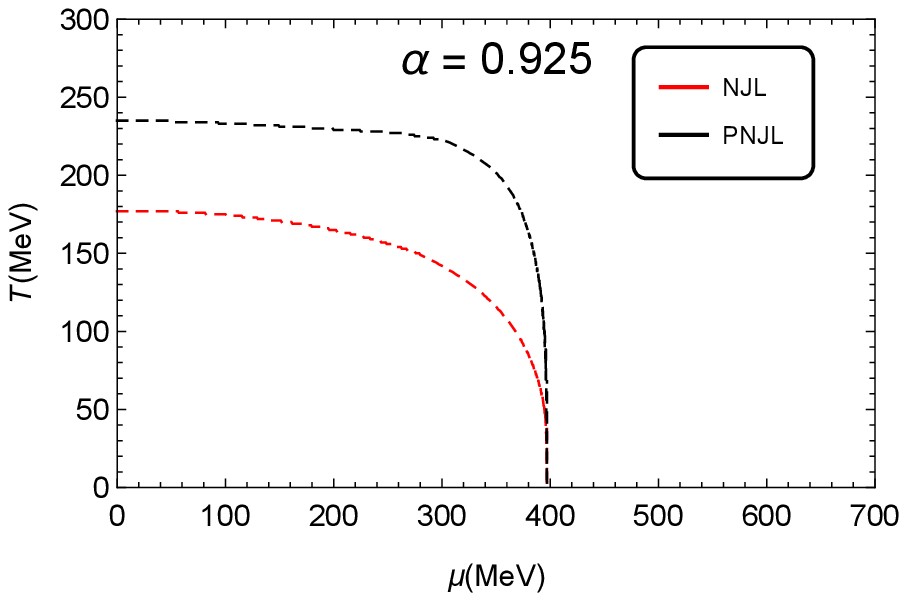}}
\hspace{0.01\textwidth}
\subfigure[]{\includegraphics[width=0.45\textwidth]{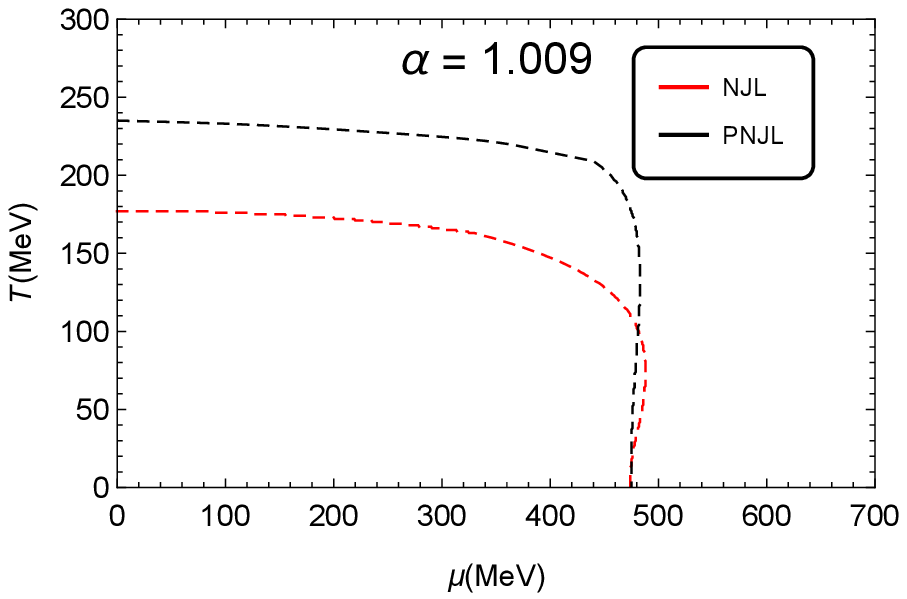}}
\subfigure[]{\includegraphics[width=0.45\textwidth]{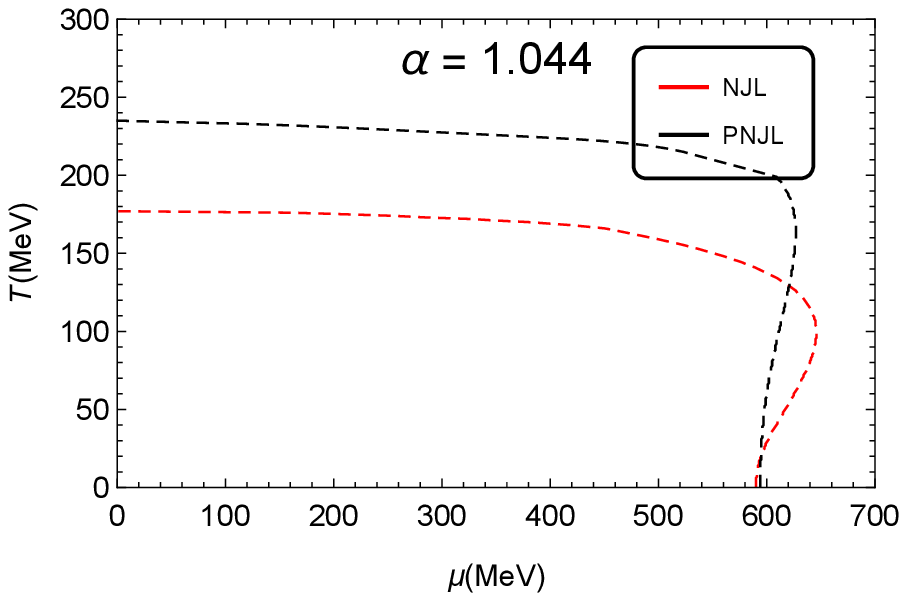}}
\caption{The comparison of chiral phase transition in the NJL and PNJL models with $\alpha = 0, 0.5, 0.925, 1.009, 1.044$. }
\label{figure2}
\end{figure}

\begin{figure}[t]
\centering
\includegraphics[width=0.45\textwidth]{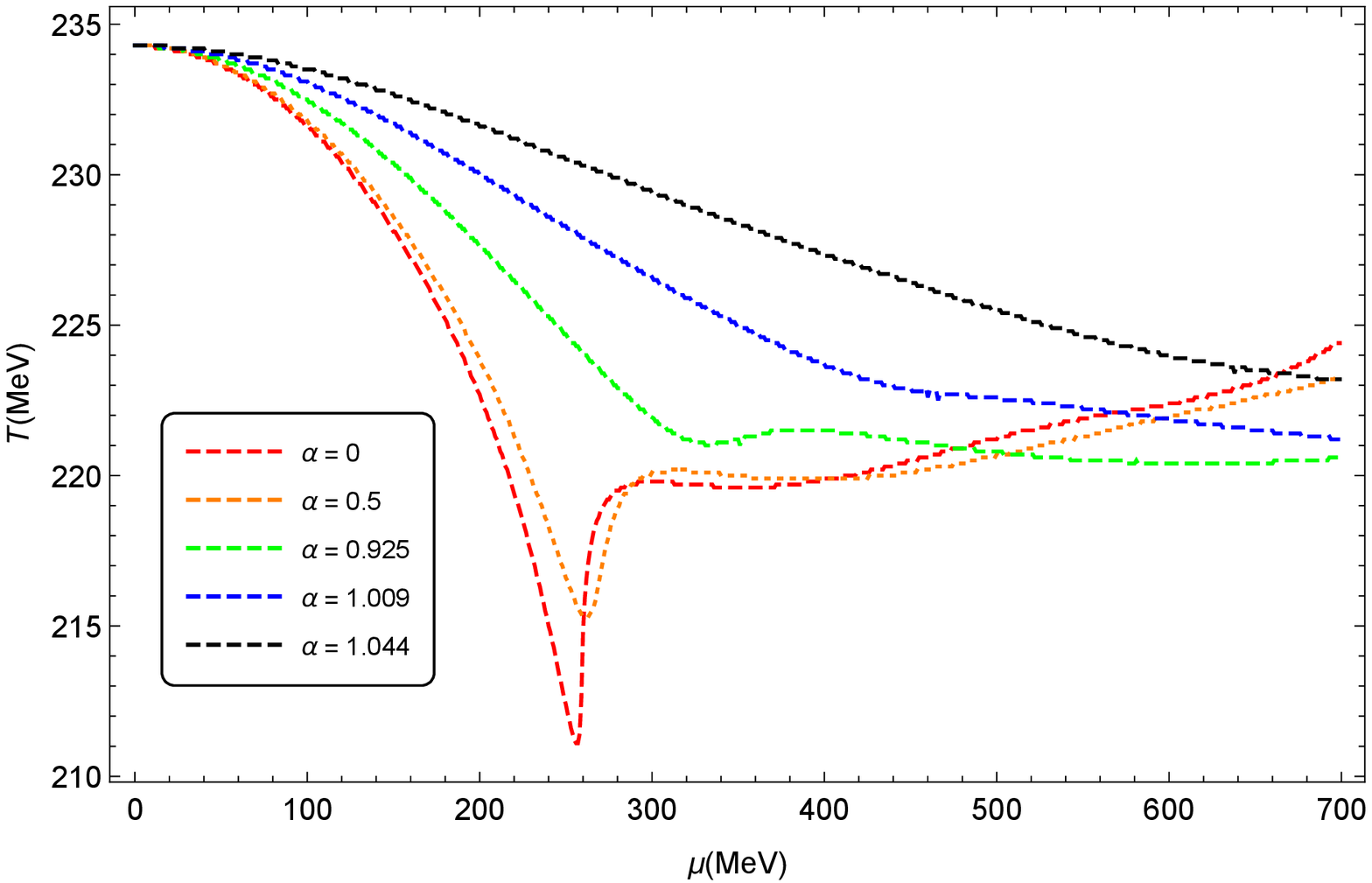}
\hspace{0.01\textwidth}
\includegraphics[width=0.45\textwidth]{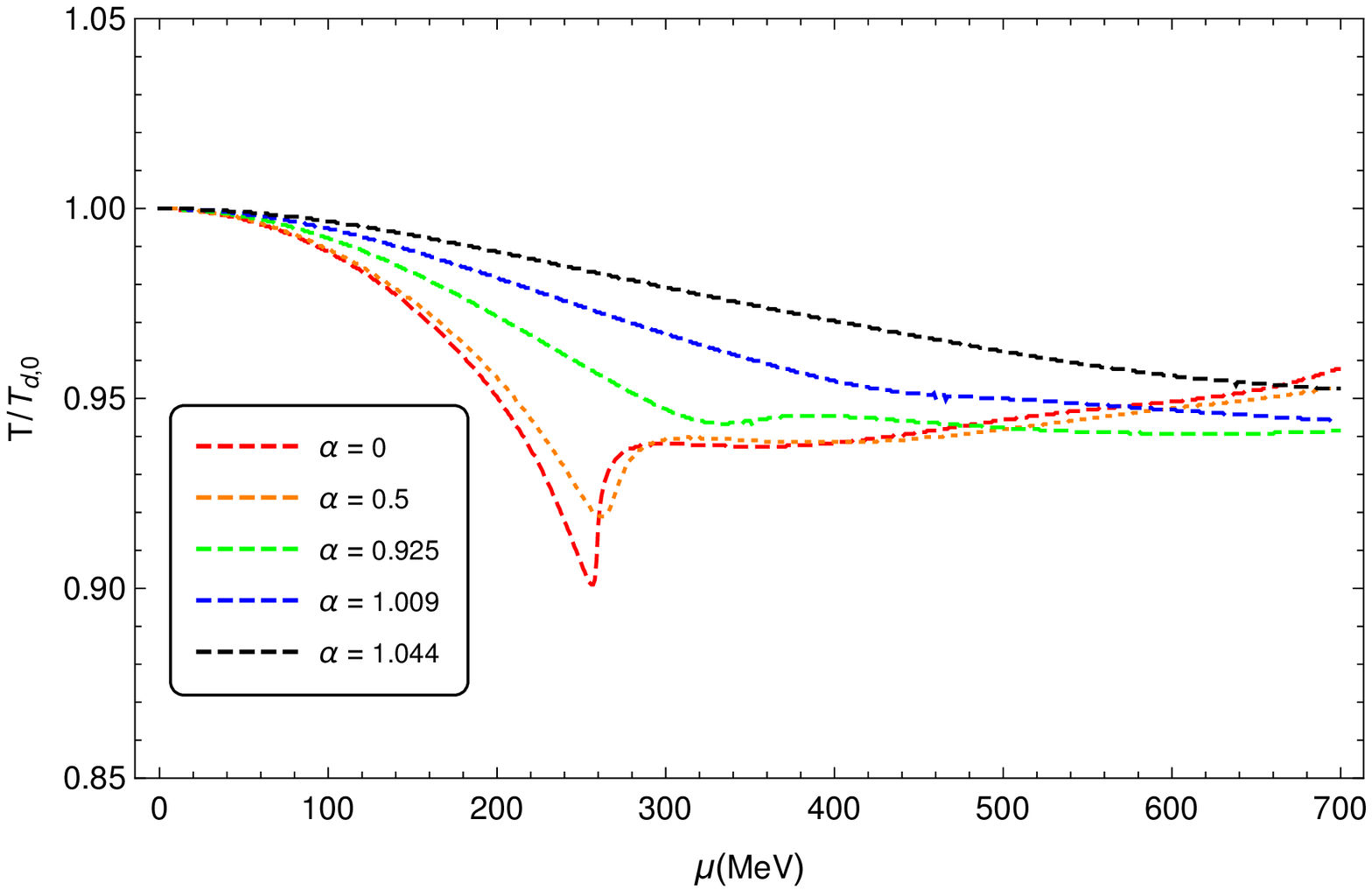}
\caption{Left panel: Deconfinement phase transition with $\alpha = 0, 0.5, 0.925, 1.009, 1.044$. Right panel: Deconfinement phase transition scaled by $T_{d,0}$ with $\alpha = 0, 0.5, 0.925, 1.009, 1.044$.}
\label{figure3}
\end{figure}

The chiral phase transition lines of the NJL and PNJL models with different values of parameter $\alpha$, which are obtained by solving Eqs.~(\ref{gap1})-(\ref{gap4}), are present in Fig.~\ref{figure1}.  We find that, when $\alpha = 0$, the CEPs are located at $(\mu_{CEP}, T_{CEP}) = (327,115)$ MeV and $(\mu_{CEP}, T_{CEP}) =  (328,47)$ MeV for the PNJL and NJL models, respectively. And when $\alpha = 0.5$, the CEPs are located at $(\mu_{CEP}, T_{CEP}) = (347,78)$ MeV and  $(\mu_{CEP}, T_{CEP}) = (348,29)$ MeV, respectively. Whereas, when $\alpha$ is large enough, e.g., $\alpha = 0.925, 1.009$ and $1.044$, there is no CEPs for both the NJL and PNJL models, and it means that the chiral phase transitions in these cases are crossover. With resort to the chiral susceptibility, we find that the critical value $\alpha_c$ is approximately 0.71 in both models, and there is first order phase transition only when $\alpha < \alpha_c$. Moreover, the critical temperature $T_c$ is around 234 MeV at zero chemical potential in the PNJL model, which is much higher than $T_c = 178$ MeV at $\mu=0$ in the NJL model. The comparison of the chiral phase transition lines between the PNJL model and the NJL model with different values of the weighting parameter $\alpha$ is given in Fig.~\ref{figure2}. The critical chemical potential $\mu_c$ at zero temperature in the NJL and PNJL models is coincided with each other at the same $\alpha$.  When $\alpha$ is smaller than 1, the phase transition line in the PNJL model is always higher than the line in the NJL model. But when $\alpha > 1$, e.g., $\alpha = 1.009$ and $\alpha = 1.044$, there is a intersection point between the two linesin the NJL and PNJL models, as shown in Fig.~\ref{figure2}.

The deconfinement phase transition lines in the PNJL model are plotted in panel(a) of  Fig.~\ref{figure3}. The critical temperatures $T_{d}$ at zero chemical potential for different values of  $\alpha$ are the same, which are almost coincided with the critical temperature of the chiral phase transition $T_c$, i.e., $T_{d}\approx T_c \approx 234$ MeV. It is found that, when $\alpha$ is small, the deconfinement phase transition lines decrease first and then increase with the increasing of the chemical potential. And as $\alpha$ increases, the kink of the curve becomes smaller and smaller, and finally it disappears. It shows that the transition line becomes smoother and smoother, and when
$\alpha$ is lager enough, the transition line decreases monotonically with the increasing chemical potential. However, from panel(b) in Fig.~\ref{figure3} we could find within a large range of chemical potential $0 - 700$ MeV, the change of the critical temperature with the increasing of the chemical potential are less than $10\%$ even for $\alpha = 0$.

\section{Summary and Discussion}
\label{sec-4}

In this paper, we studied the QCD phase transition in the framework of a self-consistent PNJL model. By comparing the chiral phase transition lines between the NJL model and the PNJL model, the Polyakov loop from the gluonic contribution helps to increase the (pseudo)critical temperature at the same chemical potential. When $\alpha>1$, there is a cross point between the phase transition lines of these two models. As expect, the incorporation of the Polyakov loop does not change the order of the chiral phase transition with the same $\alpha$. And the deconfinement phase transition lines become smoother when $\alpha$ increases. However, the lines are always relatively flatten, the change of the critical temperature are smaller than $10\%$.

As discussed in Ref.\cite{Wang:2019uwl}, the real weighting constant $\alpha$ should be determined by experiments measurement at the large chemical potential physical environments, e.g., the properties of neutron stars by observing the binary neutron star mergers~\cite{Zhao:2019xqy, Zuo:2022rks}, the meson properties from the observation of the future collision experiments, and etc. Since the gluons in the QCD matter do contribute to the equation of state, it is valuable to introduce the Polyakov loop potential to investigate the equation of state of neutron stars and the meson properties in the future.

\acknowledgements
The authors thanks to Lin Zhang for the discussion of the numerical code. The work of L.Y. is supported by the NSFC under Grant No. 11605072 and the Seeds Funding of Jilin University. The work of X.W. was supported by the start-up funding No.~4111190010 of Jiangsu University and NSFC under Grant No.~12147103.


\begin{thebibliography}{99}
\bibitem{Nambu:1961tp}
Y.~Nambu and G.~Jona-Lasinio,
Phys. Rev. \textbf{122}, 345-358 (1961)
doi:10.1103/PhysRev.122.345
\bibitem{Nambu:1961fr}
Y.~Nambu and G.~Jona-Lasinio,
Phys. Rev. \textbf{124}, 246-254 (1961)
doi:10.1103/PhysRev.124.246
\bibitem{Klevansky:1992qe}
S.~P.~Klevansky,
Rev. Mod. Phys. \textbf{64}, 649-708 (1992)
doi:10.1103/RevModPhys.64.649
\bibitem{Hatsuda:1994pi}
T.~Hatsuda and T.~Kunihiro,
Phys. Rept. \textbf{247}, 221-367 (1994)
doi:10.1016/0370-1573(94)90022-1
[arXiv:hep-ph/9401310 [hep-ph]].
\bibitem{Buballa:2003qv}
M.~Buballa,
Phys. Rept. \textbf{407}, 205-376 (2005)
doi:10.1016/j.physrep.2004.11.004
[arXiv:hep-ph/0402234 [hep-ph]].
\bibitem{Fukushima:2003fw}
  K.~Fukushima,
  Phys.\ Lett.\ B {\bf 591}, 277 (2004)
  doi:10.1016/j.physletb.2004.04.027
  [hep-ph/0310121].

\bibitem{Wang:2019uwl}
F.~Wang, Y.~Cao, Y.~Dia and H.~Zong,
Chin. Phys. C \textbf{43}, no.8, 084102 (2019)
doi:10.1088/1674-1137/43/8/084102
[arXiv:1901.05601 [nucl-th]].

\bibitem{Yu:2020dnj}
  Z.~X.~Yu, T.~Zhao and H.~S.~Zong,
  Chin.\ Phys.\ C {\bf 44}, no. 7, 074104 (2020).
  doi:10.1088/1674-1137/44/7/074104

\bibitem{Li:2018ygx}
Z.~Li, K.~Xu, X.~Wang and M.~Huang,
Eur. Phys. J. C \textbf{79}, no.3, 245 (2019)
doi:10.1140/epjc/s10052-019-6703-x
[arXiv:1801.09215 [hep-ph]].





\bibitem{Kapusta:2006pm}
J.~Kapusta and C.~Gale,
{\em Finite-temperature field theory: Principles and applications},
(Cambridge University Press, Cambridge, 2006).
\bibitem{Yang:2019lyn}
L.~K.~Yang, X.~Luo and H.~S.~Zong,
Phys. Rev. D \textbf{100}, no.9, 094012 (2019)
doi:10.1103/PhysRevD.100.094012
[arXiv:1910.13185 [nucl-th]].

\bibitem{Xue:2021ldz}
K.~Xue, X.~Yu and X.~Wang,
Chin. Phys. C \textbf{46}, no.5, 013103 (2022)
doi:10.1088/1674-1137/ac2ed3
[arXiv:2105.14323 [hep-ph]].

\bibitem{Zhao:2019xqy}
T.~Zhao, W.~Zheng, F.~Wang, C.~M.~Li, Y.~Yan, Y.~F.~Huang and H.~S.~Zong,
Phys. Rev. D \textbf{100}, no.4, 043018 (2019)
doi:10.1103/PhysRevD.100.043018
[arXiv:1904.09744 [nucl-th]].
\bibitem{Zuo:2022rks}
B.~J.~Zuo, Y.~F.~Huang and H.~T.~Feng,
Phys. Rev. D \textbf{105}, no.7, 074011 (2022)
doi:10.1103/PhysRevD.105.074011
[arXiv:2203.11791 [nucl-th]].
\end{thebibliography}
\end{document}